\documentclass[12pt,aps,preprint,nofootinbib,superscriptaddress]{revtex4}

\usepackage{epsfig}
\usepackage{axodraw}
\usepackage{amssymb}
\usepackage{amsmath}
\usepackage{amsfonts}
\usepackage{graphics}
\usepackage{cancel}
\usepackage{bbm}
\usepackage{mathrsfs}
\usepackage{slashed}

\newcommand{\Refs}{Refs.}
\newcommand{\Ref}{Ref.}
\newcommand{\Sec}{Sec.}
\newcommand{\Secs}{Secs.}

\newcommand{\eq}{Eq.}
\newcommand{\eqs}{Eqs.}
\newcommand{\fig}{Fig.}
\newcommand{\Fig}{Fig.}

\newcommand{\bld}[1]{\boldsymbol{#1}}
\newcommand{\bea}{\begin{eqnarray}}
\newcommand{\eea}{\end{eqnarray}}

\newcommand{\be}{\begin{equation}}
\newcommand{\ee}{\end{equation}}

\newcommand{\ba}{\begin{array}}
\newcommand{\ea}{\end{array}}
\newcommand{\znbbeq}{0\nu\beta\beta}
\newcommand{\znbb}{$\znbbeq$}

\newcommand{\ie}{\emph{i.e.}}
\newcommand{\eg}{\emph{e.g.}}
\newcommand{\cf}{\emph{c.f.}}

\newcommand{\hc}{\mathrm{H.c.}}

\DeclareMathOperator{\diag}{diag}

\newcommand{\type}[1]{type-#1}
\newcommand{\Type}[1]{Type-#1}

\allowdisplaybreaks

\begin{document}
\title{Neutrinoless double beta decay in seesaw models}

\author{Mattias Blennow}
\email[]{blennow@mppmu.mpg.de}
\affiliation{Max-Planck-Institut f\"ur Physik
(Werner-Heisenberg-Institut), F\"ohringer Ring 6, 80805 M\"unchen,
Germany}

\author{Enrique Fernandez-Martinez}
\email[]{enfmarti@mppmu.mpg.de}
\affiliation{Max-Planck-Institut f\"ur Physik
(Werner-Heisenberg-Institut), F\"ohringer Ring 6, 80805 M\"unchen,
Germany}

\author{Jacobo Lopez-Pavon}
\email[]{jacobo.lopez@uam.es}
\affiliation{Departamento de F\'isica Te\'orica, Universidad Aut\'onoma de Madrid, \\ 
Cantoblanco, 28049 Madrid, Spain}
\affiliation{Instituto F\'{\i}sica Te\'orica UAM/CSIC, Cantoblanco, 28049 Madrid, Spain}

\author{Javier Men\'endez}
\email[]{javier.menendez@physik.tu-darmstadt.de}
\affiliation{Departamento de F\'isica Te\'orica, Universidad Aut\'onoma de Madrid, \\ 
Cantoblanco, 28049 Madrid, Spain}
\affiliation{Instituto F\'{\i}sica Te\'orica UAM/CSIC, Cantoblanco, 28049 Madrid, Spain}
\affiliation{Institut f\"ur Kernphysik, Technische Universit\"at Darmstadt, % \\
64289 Darmstadt, Germany}
\affiliation{ExtreMe Matter Institute EMMI, GSI Helmholtzzentrum f\"ur Schwerionenforschung GmbH, \\
64291 Darmstadt, Germany
}

%\date{May 18, 2010}

\begin{abstract}
We study the general phenomenology of neutrinoless double beta decay in seesaw models.
In particular, we focus on the dependence of the neutrinoless double beta decay rate
on the mass of the extra states introduced to account for the Majorana masses of light neutrinos.
For this purpose, we compute the nuclear matrix elements as functions of the mass of the mediating fermions and estimate the associated uncertainties.
We then discuss what can be inferred on the seesaw model parameters in the different mass regimes
and clarify how the contribution of the light neutrinos should always be taken into account when deriving bounds on the extra parameters.
Conversely, the extra states can also have a significant impact,
canceling the Standard Model neutrino contribution for masses lighter than the nuclear scale
and leading to unobservable neutrinoless double beta decay amplitudes even if neutrinos are Majorana particles.
In particular, the decay rate is reduced by at least six orders of magnitude for masses of the extra states below 1~MeV in absence of extra contributions.
We also discuss how seesaw models could reconcile large rates of neutrinoless double beta decay with more stringent cosmological bounds on neutrino masses.
\end{abstract}

%\keywords{}

\preprint{MPP-2010-40}
\preprint{IFT-UAM/CSIC-10-26}
\preprint{FTUAM-10-06}
\preprint{EURONU-WP6-10-18}

\maketitle

\section{Introduction}

At present, neutrino oscillations, together with cosmological 
evidence for dark matter and dark energy, present the only 
evidence for physics beyond the Standard Model (SM). Thus, models accommodating 
neutrino masses become an important component in the search for new 
physics and it is therefore fundamental to seek experimental 
answers to questions such as whether neutrinos are Dirac or 
Majorana fermions. In this context, one of the most promising types of 
experiments is that of searching for neutrinoless double beta decay
(\znbb\ decay), in which a peak at the endpoint of the beta radiation energy is 
searched for,
see~\cite{Haxton:1985am,Doi:1985dx,Tomoda:1990rs,Suhonen:1998ck,Faessler:1999zg,Vergados:2002pv,Avignone:2007fu,Vogel:2008sx,Bilenky:2010kd} for reviews.
Since this process is lepton number violating, its observation would imply that neutrinos are Majorana fermions~\cite{Schechter:1981bd}.
Present bounds on \znbb\ decay for different nuclei have been set by the
Heidelberg-Moscow~\cite{KlapdorKleingrothaus:2000sn}, IGEX~\cite{IGEX}, CUORICINO~\cite{Arnaboldi:2008ds}, NEMO~\cite{NEMO, NEMO2}, ELEGANT~\cite{ELEGANT},
Solotvina~\cite{Solotvina} and DAMA~\cite{DAMA} collaborations, as well as geochemical measurements~\cite{Geochemical}.
In the next few years, experiments like GERDA~\cite{GERDA}, EXO~\cite{EXO}, SNO+~\cite{SNO} and CUORE~\cite{CUORE} will search for a \znbb\ decay signal with unprecedented sensitivity.
Furthermore, there are several other experiments proposed for the upcoming future, which include
CANDLES~\cite{CANDLES},
Super-NEMO~\cite{SUPERNEMO},
MAJORANA~\cite{MAJORANA},
NEXT~\cite{NEXT},
CARVEL~\cite{CARVEL},
COBRA~\cite{COBRA},
DCBA~\cite{DCBA},
MOON~\cite{MOON} and
XMASS~\cite{XMASS}.

Apart from the ongoing searches, there is a claim for a \znbb\ decay signal 
from (part of) the Heidelberg-Moscow collaboration \cite{KlapdorKleingrothaus:2001ke,KlapdorKleingrothaus:2006ff}.
However, this measurement is in apparent conflict with other bounds on neutrino masses, in particular 
those coming from the influence of neutrino masses in cosmology \cite{Hannestad:2010yi,Komatsu:2010fb}.

Among the more popular models for neutrino masses, we find the different 
types of seesaw mechanisms. In the \type{I} seesaw \cite{Minkowski:1977sc,Yanagida:1979as,Mohapatra:1979ia,GellMann:1980vs}, the SM is 
extended by the addition of extra fermion singlets. The Majorana masses of 
such singlets do not violate gauge invariance and are presently not constrained. However, a very heavy mass is usually assumed 
and a Yukawa coupling between the singlets and the SM 
neutrino fields of the same order of magnitude as the SM Yukawas is introduced. Such a 
scenario provides a natural realization of small neutrino masses even if 
other possibilities, \eg, approximate lepton number conservation \cite{Bernabeu:1987gr,Branco:1988ex}, exist.

In the current literature on \znbb\ decay, it is common to study the effects of the SM neutrinos%
\footnote{By SM neutrinos, we here mean the mass states which are predominantly composed of the SM flavor fields. Although these states can contain some admixture 
of the extra states introduced to account for neutrino masses, we will use this nomenclature for simplicity also in the remainder of the paper.}
or the extra states introduced to account for their masses independently (see, \eg, \Refs~\cite{Benes:2005hn,Atre:2009rg,Belanger:1995nh,Simkovic:1999re,delAguila:2008cj}). However, the interplay between both contributions, when combined, displays interesting 
phenomenology such as cancellations in certain regimes that is otherwise lost. By considering these contributions independent, the presence of some extra component to the Majorana neutrino mass different from the extra states and some degree of fine tuning is implicitly assumed. 

In the present work, we will compute the nuclear matrix element (NME) involved in the \znbb\ decay rate
without any assumption on the neutrino mass mediating the process, describing in detail the approximations involved.
The uncertainties associated to each of these approximations are also discussed, so as to estimate the total uncertainty on the final NME. The results of this computation are publicly available in Appendix~\ref{nme_table}. %at \Ref~\cite{webpage}.
We will also discuss in detail the interplay between the contributions of the SM neutrinos and the extra states in seesaw models and discuss under which conditions they can be considered independent. In doing so we will cover the full parameter 
space and deduce what implications can actually be inferred on the 
models from observations.
Finally, we will also comment on how the contribution of extra states with different mass scales 
could allow for a large \znbb\ decay rate even in presence of more stringent bounds on neutrino masses, such as those derived from cosmology. In particular, it will be shown that the Heidelberg-Moscow claim requires a tuning of only about 50~\% amongst the extra contributions in order to be compatible with the present cosmology bounds.

The remainder of this work is organized as follows: In \Sec~\ref{sec:NMEcomp},
we review the physics and assumptions used to compute the NMEs and perform these calculations in order to get their values as
functions of the mass of the exchanged fermions.
Next, in \Sec~\ref{sec:pheno}, we discuss the general phenomenology of the \znbb\ process and the approximations that are usually made in \znbb\ decay analyses.
The realizations of the \znbb\ decay signal in the different types of seesaw models are treated in \Sec~\ref{Sec:Models}
before we summarize and give our conclusions in \Sec~\ref{sec:summary}.

\section{Computation of nuclear matrix elements}
\label{sec:NMEcomp}

\subsection[]{Obtaining the rate of $\bld{\znbbeq}$ decay \label{Sec:NME}}

The \znbb\ decay is mediated by the weak Hamiltonian:
\begin{equation}
H_{W}=\frac{G_F}{\sqrt{2}}\left(j_{L\mu}J_{L}^{\mu\dagger}\right)+ \hc ,
\label{eq:H_w}
\end{equation}
where $j_{L\mu}$ is the leptonic current,
which consists of the electron coupled to left handed electron neutrino.
The neutrino can be written as a linear combination of the light and heavy mass eigenstates $\nu_i$ given by
the mixing matrix $U$:
\begin{equation}
j_{L\mu}=\overline{e}\gamma_{\mu}\left(1-\gamma_{5}\right)\nu_{eL}, \qquad \nu_{eL}=\sum_{i}U_{ei}\nu_{iL}.
\label{eq:j_mu}
\end{equation}
On the other hand, the hadronic (nuclear) current $J_{L}^{\mu}$ can be obtained phenomenologically by imposing symmetry requirements
to the more general combination that can be built with the available Lorentz vectors
$p_{n}^{\mu}$, $p_{p}^{\mu}$ and $\gamma^{\mu}$, the neutron and proton four-momenta and the spin matrices, respectively.
We also need to assume the impulse approximation, \ie, that nucleons in nuclei can be treated as free when dealing with the weak interaction.
Then imposing Lorentz, parity and time-reversal invariance the nuclear current is given by\footnote{
Second-class currents $g_{S}(p^{2})p^{\mu}$ and $g_{T}(p^{2})\frac{\sigma^{\mu\nu}}{2m_{N}}p_{\nu}\gamma_{5}$, for which there is no experimental evidence, will be ignored.}
\begin{equation}
J_{L}^{\mu\dagger} = \overline{\Psi}\tau^{-}
\left[\frac{}{}g_{V}(p^{2})\gamma^{\mu}
+ig_{M}(p^{2})\frac{\sigma^{\mu\nu}}{2m_{N}}p_{\nu}
-g_{A}(p^{2})\gamma^{\mu}\gamma_{5}
-g_{P}(p^{2})p^{\mu}\gamma_{5}
\right]\Psi,
\label{eq:J_mu}
\end{equation}
\noindent
where $p^{\mu}=p_{n}^{\mu}-p_{p}^{\mu}$ is the transferred momentum from hadrons to leptons, $m_N$ is the nucleon mass, $\Psi$ represents a nucleon field
and $\tau^{-}$ is the isospin lowering operator, \ie, it turns a neutron into a proton.

The form factors, $g_{V}$, $g_{M}$, $g_{A}$ and $g_{P}$, are real functions of the Lorentz scalar $p^2$.
Their values at zero-momentum transfer are known as the vector, magnetic, axial and  pseudoscalar coupling constants, respectively.
Note that in single $\beta$ and two-neutrino $\beta\beta$ decays only the vector and axial terms are usually considered, due to the small transferred momenta ($\lesssim1$ MeV).
The magnetic and pseudoscalar couplings can be written in terms of the vector and axial ones by assuming the conserved vector current (CVC)
and the partially conserved axial current (PCAC) hypotheses \cite{Towner:1995za}.
The CVC hypothesis also implies that $g_{V}(0)=1$ in the nuclear medium.

We now take the non relativistic approximation to the hadronic current.
If terms are kept up to $|\mathbf{p}|/m_N$ ($|\mathbf{p}|\simeq 100$~MeV as will be discussed below) 
and the energy transfer between nucleons is neglected ($E\simeq\mathbf{p}^2/2m_N$), we are left with\footnote{
Nuclear recoil terms also come at first order in $1/m_N$, being proportional to $\mathbf{p}_p+\mathbf{p}_n$ instead of $\mathbf{p}$.
However, their leading contribution is suppressed an extra order of magnitude because of their odd-parity character, which requires electron $p$-waves.
Hence these terms will be neglected. Their contribution will be discussed when referring to the electron $s$-wave approximation.}
\begin{equation}
J_{L}^{\mu\dagger}(\mathbf{x})=\sum_{n=1}^{A}\tau_{n}^{-}
\left[g^{\mu0}J_{n}^{0}(p^{2})
+g^{\mu k}J_{n}^{k}(p^{2})\right]
\delta(\mathbf{x}-\mathbf{r}_n),
\label{eq:J_norel}
\end{equation}
\noindent
where
\begin{eqnarray}
J_{n}^{0}(p^{2}) & =& g_{V}(p^{2}),\nonumber \\
\mathbf{J}_{n}(p^{2}) & = &
ig_{M}(p^{2})\frac{\bld{\sigma}_{n}\times\mathbf{p}}{2m_{N}}+
g_{A}(p^{2})\bld{\sigma}_{n}-
g_{P}(p^{2})\frac{\mathbf{p}\left(\mathbf{p}\cdot\bld{\sigma}_{n}\right)}{2m_{N}}.
\label{eq:Jota}
\end{eqnarray}
Hence, we have a sum over all $A$ nucleons of the nucleus, whose coordinates are denoted by $\mathbf{r}_{n}$.
Note that nucleon operators present in the nucleon fields from now on will be included in the nuclear wavefunctions.

The momentum dependence of the couplings is usually parametrized by the standard dipolar form \cite{Vergados:1982wr},
and takes into account that nucleons are not point particles but finite size bodies, \ie, the nucleon structure.
Since in \znbb\ decay the neutrino is being exchanged in t-channel and the outgoing electrons have essentially the same energy,
the energy exchange can be neglected and thus $p^2 \simeq - \mathbf{p}^2$.
Then, the form factors look like
\begin{align}
g_{V}(p^{2}) & = \frac{g_{V}(0)}{\left(1+\frac{\mathbf{p}^{2}}{\Lambda_{V}^{2}}\right)^{2}},
\qquad
g_{M}(p^{2})
=\left(\mu_{p}-\mu_{n}\right)g_{V}(p^{2}), \nonumber \\
g_{A}(p^{2})
& =\frac{g_{A}(0)}{\left(1+\frac{\mathbf{p}^{2}}{\Lambda_{A}^{2}}\right)^{2}},
\qquad
g_{P}(p^{2}) = \frac{2m_{N}g_{A}(p^{2})}{\left(\mathbf{p}^{2}+m_{\pi}^{2}\right)},
\label{eq:g_q}
\end{align}
\noindent
where $m_\pi$ is the pion mass and $\mu_p$ and $\mu_n$ denote the proton and neutron anomalous magnetic moments, respectively.
The values of the cutoffs of the vector and axial nucleon form factors, $\Lambda_{V}=0.85$ GeV and $\Lambda_{A}=1.09$ GeV,
are taken from experimental observations~\cite{Dumbrajs:1983jd,Ahrens:1988rr}.
Their effect is to weaken the couplings for large momentum transfers,
\ie, they avoid contributions arising from nucleons being too close to one another.
These form factors are commonly denoted as the finite nuclear size (FNS) terms.

With this Hamiltonian the rate of the  \znbb\ decay can be calculated by using the second order Fermi's Golden Rule~\cite{Doi:1985dx}:
\begin{equation}
d\Gamma_{0\nu\beta\beta}=2\pi\sum_{\rm spin} \left|R_{0\nu\beta\beta}\right|^{2}\delta(\varepsilon_{1}+\varepsilon_{2}+E_{f}-E_{i})d\Omega_{e_{1}}d\Omega_{e_{2}},
\label{eq:anch_onu}
\end{equation}
\noindent
where the transition amplitude is given by
\begin{eqnarray}
R_{0\nu\beta\beta} & = &\left(\frac{G_{F}}{\sqrt{2}}\right)^{2}
\int d\mathbf{x}\int d\mathbf{y}\frac{1}{\sqrt{2}}\left(1-P_{12}\right) \times  \nonumber \\
&&\sum_{a,j} 
\frac{\left\langle N_{f};e_1,e_2\right|J_{L}^{\mu\dagger}(\mathbf{x})j_{L\mu}(\mathbf{x})\left|N_{a};e_1,\nu_j\right\rangle
\left\langle N_{a};e_1,\nu_j\right|J_{L}^{\rho\dagger}(\mathbf{y})j_{L\rho}(\mathbf{y})\left|N_{i}\right\rangle}
{\omega_j+E_a-\left(E_i+\varepsilon_1\right)}.
\label{eq:Ronu}
\end{eqnarray}
\noindent
The operator $P_{12}$ is included to fulfill antisymmetry for the electrons,
whose energies are denoted by $\varepsilon_1$, $\varepsilon_2$.
The energy of the virtual neutrino $\nu_j$ is denoted by $\omega_j$, and
$E_a$ is the energy of the virtual intermediate nuclear state $\left|N_{a}\right\rangle$.

The leptonic part of the numerator in \eq~(\ref{eq:Ronu}) can be written more explicitly as
\begin{align}
- i\sum_{j} & U_{ej}^{2}
\overline{e}(x)\gamma_\mu\left(1-\gamma_5\right)
\int \frac{d^4p}{\left(2\pi\right)^{4}}\frac{e^{ip\cdot(x-y)}}{p^2-m_j^2}
\left(\slashed{p}+m_j\right)C^T\left(1-\gamma_5\right)\gamma_\rho \overline{e}^{T}(y) \nonumber \\
= & -i \sum_{j}U_{ej}^{2}m_j
\int \frac{d^4p}{\left(2\pi\right)^{4}}\frac{e^{ip\cdot(x-y)}}{p^2-m_j^2}
2\overline{e}(x)\gamma_\mu\left(1-\gamma_5\right)\gamma_\rho {e}^{C}(y)\nonumber \\
= &-i \sum_{j}U_{ej}^{2}m_j
\int \frac{d \mathbf{p}}{\left(2\pi\right)^{3}}\frac{e^{i\mathbf{p}\cdot(\mathbf{x}-\mathbf{y})}}{\omega_j}
\overline{e}(x)\gamma_\mu\left(1-\gamma_5\right)\gamma_\rho e^{C}(y).
\label{eq:Contraction}
\end{align}
\noindent
This term turns out to be proportional to the neutrino masses $m_j$ because of the left handed character of both leptonic currents.
The transition amplitude is then
\begin{align}
R_{0\nu\beta\beta} = \frac{G_{F}^2}{\sqrt{2}}&\frac{g_{A}^{2}(0)m_e}{8\pi R}
\int d\mathbf{x}\int d\mathbf{y} \left(1-P_{12}\right)\overline{e}(\varepsilon_{1},\mathbf{x})\gamma_{\mu}
\left(1-\gamma_{5}\right)\gamma_{\rho}e^{C}(\varepsilon_{2},\mathbf{y})\times \nonumber \\
&\sum_{j}\frac{U_{ej}^{2}m_j}{m_e}\frac{R}{g_{A}^{2}\left(0\right)}
\int \frac{d \mathbf{p}}{2\pi^{2}}\frac{e^{i\mathbf{p}\cdot(\mathbf{x}-\mathbf{y})}}{\omega_j}
\sum_{a} \frac{\left\langle N_{f}\right|J_{L}^{\mu\dagger}(\mathbf{x})\left|N_{a}\right\rangle
\left\langle N_{a}\right|J_{L}^{\rho\dagger}(\mathbf{y})\left|N_{i}\right\rangle}
{\omega_j+\mu_{a}-\frac{1}{2}\left(\varepsilon_{1}-\varepsilon_{2}\right)}.
\label{eq:Ronu_2}
\end{align}
\noindent
The axial coupling $g_{A}^{2}(0)$, the electron mass $m_e$ and the nuclear radius $R$
have been introduced for convenience and to make the second line in \eq~(\ref{eq:Ronu_2}) dimensionless.
For the nuclear radius we have taken $R=1.2A^{1/3}$~fm.
Using energy conservation, we have also rewritten the denominator using the new parameter
\begin{equation}
\mu_{a}\equiv E_{a}^{m}-\frac{1}{2}\left(E_{i}+E_{f}\right),
\label{eq:mu_a}
\end{equation}
which gives the relative energy of the (virtual)
state of the intermediate nucleus with respect to the mean energy of the initial and final states.
Typical values of this parameter for the different decays are $\sim10$~MeV~\cite{Haxton:1985am}.

In the following, two approximations will be made:
\begin{itemize}
	\item Closure approximation.
	\item Approximation of $0^+$ final states and electrons emitted in $s$-wave.
\end{itemize}
The first of these takes advantage of the high momentum of the virtual neutrino $\left|\mathbf{p}\right|\simeq 100$~MeV.
This nuclear scale comes from the integral over the transferred momentum of \eq~(\ref{eq:Ronu_2}).
For light neutrinos, the integrand of this equation is approximately proportional to $\left|\mathbf{p}\right|/(\left|\mathbf{p}\right|+\mu_a)$ in radial coordinates.
Hence, momenta $\lesssim$~10~MeV will be disfavoured in the transition.
In addition, if we recall the form of the FNS terms in \eq~(\ref{eq:g_q}),
we see that transferred momenta above the cutoffs $\simeq 1$ GeV will be suppressed as well.
On the other hand, the leading contribution of the exponential term will arise when $\mathbf{p}\cdot(\mathbf{x}-\mathbf{y})\simeq 1$.
Since nucleons are typically few fermis (femtometers) apart in nuclei, the nuclear wavefunctions will select
the preferred momentum for the virtual neutrino to be $\left|\mathbf{p}\right|\simeq 100$~MeV.
Thus, we would expect it to be the typical virtual neutrino momentum of the decay.
This result has been confirmed by explicit calculation~\cite{Simkovic:2007vu,Menendez:2009dis}.

In the case of heavy neutrinos,
the transition operator will now have stronger preference for larger momenta $\left|\mathbf{p}\right|\gtrsim m_j$,
since $\omega_j=\sqrt{m_j^2+\mathbf{p}^2}$ appears in the denominator in \eq~(\ref{eq:Ronu_2}).
Therefore, the tendency of the nuclear interaction for $\left|\mathbf{p}\right|\simeq 100$~MeV can be overcome resulting
in large transferred momenta that would imply internucleonic distances much shorter than $\sim0.1$~fm.
Again, such a distance is very suppressed by taking the FNS terms into account, \ie, the nucleon structure information.
Hence, in these cases, the reduction due to the FNS
effects is very large so that we end up with a $\left|\mathbf{p}\right|$ value of a few hundreds of MeV at most,
which is also the expected value for a process taking place between nucleons in nuclei.

In any case, the term $\left(\varepsilon_{1}-\varepsilon_{2}\right)$,
which can amount up to a couple of MeV and vanishes on average, can be safely neglected.
Moreover, the intermediate state energies $E_{a}$, which can differ
from one another by a few MeV, can also be replaced by an average value
$\left\langle E^m\right\rangle$. Thus, only a common parameter
\begin{equation}
\mu_{a}\simeq\mu\equiv\left\langle E^{m}\right\rangle -\frac{1}{2}\left(E_{i}+E_{f}\right)
\label{eq:<mu>}
\end{equation}
is required~\cite{Haxton:1985am}.
With the removal of the dependence on the actual energy of the intermediate states,
it follows that they are no longer needed in the calculation, since
the closure relation can be applied. Thus, the hadronic part in \eq~(\ref{eq:Ronu_2}) is now
\begin{align}
\sum_{a}
\frac{\left\langle N_{f}\right|J_{L}^{\mu\dagger}(\mathbf{x})\left|N_{a}\right\rangle
\left\langle N_{a}\right|J_{L}^{\rho\dagger}(\mathbf{y})\left|N_{i}\right\rangle}
{\omega_j+\mu_{a}-\frac{1}{2}\left(\varepsilon_{1}-\varepsilon_{2}\right)}
&\simeq\frac{1}{\omega_j+\mu}\sum_{a}
\left\langle N_{f}\right|J_{L}^{\mu\dagger}(\mathbf{x})\left|N_{a}\right\rangle
\left\langle N_{a}\right|J_{L}^{\rho\dagger}(\mathbf{y})\left|N_{i}\right\rangle \nonumber \\
 &=\frac{1}{\omega_j+\mu}
\left\langle N_{f}\right|J_{L}^{\mu\dagger}(\mathbf{x}) J_{L}^{\rho\dagger}(\mathbf{y})\left|N_{i}\right\rangle.
\end{align}
This closure approximation has been shown to be correct to more than 90~\%,
using the quasiparticle random phase approximation method \cite{Muto:1994hi}, to be presented in Sec. \ref{Sec:wf}.

As for the limitation of our study to transitions to $0^{+}$ final states,
and to cases where electrons are emitted in $s$-wave, corrections are expected to be of the order of 1~\% at most.
In the case of $p$-waves, they are suppressed to $s$-waves by an order of magnitude at least \cite{Doi:1985dx}.
Moreover, since they have odd-parity, they need odd-parity terms in the current to couple to a $J^{+}$ final state and these only appear at $\mathcal{O}\left( \left|\mathbf{p}\right|/m_N \right)$
, see note on \eq~(\ref{eq:J_norel}).
Thus, in the end we have a contribution of the order of 1~\% of the leading ones.
On the other hand, since all the final nuclei of \znbb\ decay have even number of both protons and neutrons, their ground states are always $0^{+}$.
Any transition to an excited final state will be suppressed by a phase space factor, which in this case is approximately proportional to $Q_{\beta\beta}^5$,
where $Q_{\beta\beta}$ is the energy available for the decay.
Considering this factor, the only other low-lying final states of interest are $2^{+}$ excited states. 
Moreover, apart from the phase space suppression, these final states also need electron $p$-waves due to angular momentum coupling, since two electron $s$-waves can only couple to angular momentum 0 or 1.
Hence, these transitions can also be safely neglected.

Within these approximations, the transition amplitude can be written as
\begin{align}
R_{0\nu\beta\beta} =&\frac{G_{F}^{2}}{\sqrt{2}}\frac{g_{A}^{2}\left(0
\right)m_e}{4\pi R}
\overline{e}(\varepsilon_{1})\left(1
+\gamma_{5}\right)e^{C}(\varepsilon_{2}) \times \nonumber \\
& \sum_{j}U_{ej}^{2}\frac{m_j}{m_e} 
\left\langle 0^{+}_{f}\right|\sum_{n,m}\tau_{n}^{-}\tau_{m}^{-}
\frac{R}{g_{A}^{2}\left(0\right)} \int \frac{d \mathbf{p}}{2\pi^{2}}e^{i
\mathbf{p}\cdot(\mathbf{r}_n-\mathbf{r}_m)}
\frac{\Omega_{nm}(\mathbf{p}^2)}
{\omega_j\left(\omega_j+\mu\right)}\left|0^{+}_{i}\right\rangle,
\label{eq:Ronu_3}
\end{align}
\noindent
with the tensor operator $\mathbf{S}_{nm}^{\mathbf{p}}=3\left(\mathbf{\hat{p}}\cdot\bld{\sigma}_{n})(\mathbf{\hat{p}}\cdot\bld{\sigma}_{m}\right)-\bld{\sigma}_{n}\cdot\bld{\sigma}_{m}$.
The functions $h(\mathbf{p}^2)$ can be labeled according to the terms of the hadronic current [see \eq~(\ref{eq:J_mu})]
from which they originate:
\begin{align}
h^{F}(\mathbf{p}^2) & = h_{VV}^{F}(\mathbf{p}^2),\nonumber \\
h^{GT}(\mathbf{p}^2) & = h_{AA}^{GT}(\mathbf{p}^2)+h_{AP}^{GT}(\mathbf{p}^2)+
h_{PP}^{GT}(\mathbf{p}^2)+h_{MM}^{GT}(\mathbf{p}^2),\nonumber \\
h^{T}(\mathbf{p}^2) & = h_{AP}^{T}(\mathbf{p}^2)+h_{PP}^{T}(\mathbf{p}^2)+h_{MM}^{T}(\mathbf{p}^2).
\label{eq:h_xx}
\end{align}
\noindent
Their explicit form can be found in \Ref~\cite{Simkovic:1999re}.
It was in this work that the importance of the non leading terms
[\ie, all but $h_{AA}^{GT}$ and $h_{VV}^{F}$ in Eq.~(\ref{eq:h_xx})] was first shown.
They are referred to as higher order components of the nuclear current (HOC).
Notice that, since these terms are of orders ${\left|\mathbf{p}\right|}/{m_N}$ and $\left({\left|\mathbf{p}\right|}/{m_N}\right)^2$
in the current,\footnote{
The terms labeled $h_{PP}$ and $h_{MM}$ come at order $\left({\left|\mathbf{p}\right|}/{m_N}\right)^2$.
The reason to keep these second order terms is the enhancement of the coupling constants $g_P$ and $g_M$ due to the factors
$2m_N\left|\mathbf{p}\right|/\left(\mathbf{p}^{2}+m_{\pi}^{2}\right)\simeq7$ and $\left(\mu_{p}-\mu_{n}\right)=3.70$, respectively, see \eq~(\ref{eq:g_q}).
In each of these terms the corresponding factor appears squared.}
their contribution will be enhanced for larger transferred momentum, \ie, for heavy neutrinos.

It is easy to insert $R_{\znbbeq}$ into \eq~(\ref{eq:anch_onu}) and derive an expression for the \znbb\ decay rate:
\begin{equation}
\frac{\Gamma_{0\nu\beta\beta}}{\ln2}=
G_{01}\left|\sum_{j}U_{ej}^{2}\frac{m_j}{m_e}M^{0\nu\beta\beta}(m_j)\right|^{2}.
\label{eq:t-1_rep}
\end{equation}
Here, $G_{01}$ is a well known kinematic factor, and comes essentially from the leptonic degrees of freedom.
It can be written explicitly as
\begin{align}
G_{01}& =\frac{\left[G_{F}g_{A}(0)\right]^{4}m_{e}^{2}}{64\pi^{5} R^{2} \ln2}\int F_{0}(Z,\varepsilon_{1})F_{0}(Z,\varepsilon_{2})
q_{1}q_{2}\varepsilon_{1}\varepsilon_{2}\delta(\varepsilon_{1}+\varepsilon_{2}+E_{f}-E_{i}) d\varepsilon_{1}d\varepsilon_{2}d\left(\mathbf{\hat{q}_{1}\cdot\hat{q}_{2}}\right),
\label{eq:G01}
\end{align}
where $F_{0}(Z,\varepsilon)$ are the so-called Fermi functions, with $Z$ the proton number
and $\mathbf{q_i}$ the electron momenta.
The Fermi functions also depend on the nuclear radius $R$ and
their explicit form is
\begin{equation}
 F_0(Z,\varepsilon)=\frac{4}{\Gamma^2(2\gamma_1+1)}
\left(2qR\right)^{2\left(\gamma_1-1\right)}\left|\Gamma (\gamma_1+iy )\right|^2 e^{\pi y};
\qquad\gamma_1=\sqrt{1-\left(\alpha Z\right)^2},\quad y=\frac{\alpha Z\varepsilon}{q}.
\end{equation}
The quantity $M^{0\nu\beta\beta}(m_j)$ is the nuclear matrix element,
which takes into account the initial and final nuclear wavefunctions and the transition operator.
This operator originates from both the nuclear currents and the virtual neutrino.
The NME is given by
\begin{equation}
M^{0\nu\beta\beta}(m_j)=\left\langle 0_{f}^{+}\right|
\sum_{n,m}\tau_{n}^{-}\tau_{m}^{-}
\frac{R}{g_{A}^{2}(0)}\int \frac{d \mathbf{p}}{2\pi^{2}}e^{i\mathbf{p}\cdot(\mathbf{r}_n-\mathbf{r}_m)}
\frac{\Omega_{nm}(\mathbf{p}^2)}{\omega_j\left(\omega_j+\mu\right)}\left|0_{i}^{+}\right\rangle.
\label{eq:NME_mnu}
\end{equation}
When the integral over $\mathbf{p}$ is performed, we obtain
\begin{align}
M^{0\nu\beta\beta}(m_j) & =\left\langle 0_{f}^{+}\right|\sum_{n,m}\tau_{n}^{-}\tau_{m}^{-}
\left(-V^{F}(r)+
V^{GT}(r){\bld{\sigma}_{n}\bld{\sigma}_{m}}-
V^{T}(r)\mathbf{S}_{nm}^{r}\right)\left|0_{i}^{+}\right\rangle,
\label{eq:NME}
\end{align}
\noindent
where $r=\left|\mathbf{r}_n-\mathbf{r}_m\right|$ is the distance between the decaying neutrons
and the $V(r)$ are the so-called neutrino potentials.
Before the radial integration over $\left|\mathbf{p}\right|$ they are given by
\begin{align}
V_{x}^{F/GT}(r) & = \frac{2}{\pi}\frac{R}{g_{A}^{2}(0)}\int_{0}^{\infty}
j_{0}(\left|\mathbf{p}\right|r)\frac{h_{x}^{F/GT}(\mathbf{p}^2)}{\omega_j\left(\omega_j+\mu\right)}\mathbf{p}^2\,d\left|\mathbf{p}\right|,\nonumber \\
V_{x}^{T}(r) & = -\frac{2}{\pi}\frac{R}{g_{A}^{2}(0)}\int_{0}^{\infty}
j_{2}(\left|\mathbf{p}\right|r)\frac{h_{x}^{T}(\mathbf{p}^2)}{\omega_j\left(\omega_j+\mu\right)}\mathbf{p}^2\,d\left|\mathbf{p}\right|,
\label{eq:V_int}
\end{align}
\noindent
where $j_{n}(x)$ are the spherical Bessel functions.

\subsection{Calculation of the nuclear wavefunctions within the Interacting Shell Model \label{Sec:wf}}

As can be seen from \eq~(\ref{eq:NME}), a key ingredient in the calculation
of the NMEs are the wavefunctions of the initial and final nuclei.
This is a complicated nuclear structure problem which cannot be solved
in the complete space, \ie, taking into account all neutrons and protons
of the corresponding nucleus in all their possible configurations.
Thus, truncated valence spaces and effective interactions are used
to solve the nuclear many body problem. 
As a consequence of this, a fully consistent treatment would demand regularizing the \znbb\ decay
operator in \eq~(\ref{eq:NME}) using the same prescription as for the bare nuclear interaction~\cite{Hjorth-Jensen:1995ko}.
This has only been performed very recently \cite{Engel:2009ha,Simkovic:2009pp}.

Instead, we will simplify the problem keeping the bare \znbb\ decay operator and including new correlations [called short range correlations (SRC)] in the calculation via a general prescription.
The findings of \Refs~\cite{Engel:2009ha} and~\cite{Simkovic:2009pp} show that for light neutrinos, the effect of these correlations is rather moderate once FNS terms have been taken into account,
of the order of 5\% correction to the NME. In order to implement these SRC one needs to assume some prescription and the most commonly used
are either a Jastrow-type function \cite{Wu:1985xy} or a unitary correlation operator method (UCOM) transformation \cite{Kortelainen:2007rn}.
The NME is thus transformed as
\begin{align}
\left\langle 0_{f}^{+}\right|O^{0\nu\beta\beta}\left|0_{i}^{+}\right\rangle _{SRC}=
\left\langle 0_{f}^{+}\right|U^{\dagger}O^{0\nu\beta\beta}U\left|0_{i}^{+}\right\rangle=
\left\langle 0_{f}^{+}\right|\widetilde{O}^{0\nu\beta\beta}\left|0_{i}^{+}\right\rangle,
\label{eq:ucom_em}
\end{align}
\noindent
where $U$ is either a Jastrow-type function or a UCOM transformation.
The actual parametrizations can be found in \Refs~\cite{Simkovic:2009pp} and \cite{Roth:2005pd}, respectively.
In our calculations have used the UCOM prescription, even though similar results are expected within the Jastrow approach.
Note that these SRC terms, which do not have much importance for light neutrinos, will be more relevant for heavy ones, which require shorter distances between the decaying nucleons.

Mainly two different methods are used to obtain the NMEs for the \znbb\ decay,
the quasiparticle random phase approximation (QRPA) \cite{Suhonen:1998ck,Faessler:1999zg}
and the interacting shell model (ISM) \cite{Caurier:2004gf,Caurier:2007wq}.
The QRPA includes relatively large valence spaces but is not able to comprise
all the possible configurations. On the other hand, the ISM is limited to smaller
configuration spaces, but all possible correlations within the space can be included.

When comparing the NMEs obtained by both methods a clear disagreement is found, the ISM values
being about 1.5-2 times smaller than QRPA ones \cite{Menendez:2009dis}. This disagreement is not overcome when taking into account the estimated errors
of both calculations and it is under discussion whether the difference is mainly due to the lack of correlations of QRPA calculations
(which reduce the value of the NME), the small valence space used in the ISM (to be discussed in Section~\ref{Sec:error}) or both \cite{Caurier:2007wq,Simkovic:2007vu}.
However, it must be stressed that due to the theoretical effort made over the last years this disagreement is now much less severe than it was
five years ago, and studies using recently available experimental information for the decay of $^{76}$Ge
suggest that the present situation can be improved \cite{Simkovic:2008cu,Menendez:2009oc}.
This applies to NMEs obtained with light neutrino exchange.
In the heavy neutrino case, since the available QRPA results are rather outdated \cite{Simkovic:1999re}, it is probably not meaningful to compare them
with those of the present work. However, the same relation between ISM and QRPA results of the light neutrino case is to be expected,
since the difference between these methods lies on the calculation of the wavefunctions,
and therefore it should not be very much dependent on changes on the transition operator, for whom both methods give an equivalent description \cite{Menendez:2009dis}.

\begin{figure}
\begin{center}
\includegraphics[width=0.7\textwidth]{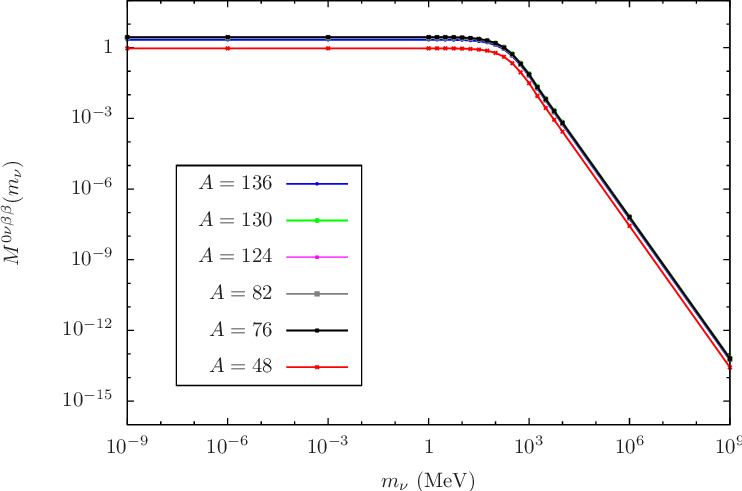}
 \bigskip 
 \caption{The nuclear matrix element dependence on the neutrino mass. The data depicted in this figure is available in Appendix~\ref{nme_table}. %at \Ref~\cite{webpage}.
 \label{fig:NME}}
\end{center}
\end{figure}

In this work, we have used ISM nuclear wavefunctions. 
Following the considerations of \Sec~\ref{Sec:NME}, we have performed calculations for
the $m_j$ dependent \znbb\ decay NMEs of the emitters $^{48}$Ca, $^{76}$Ge,
$^{82}$Se, $^{124}$Sn, $^{130}$Te and $^{136}$Xe, using the ISM
coupled code NATHAN~\cite{Caurier:2004gf}, ideally adapted for the calculation of $0^{+}$ states.
Full diagonalizations are accomplished within different valence spaces
and effective interactions. For instance, the decay of $^{48}$Ca
is studied in the $pf$ major shell, where the KB3 interaction \cite{Poves:1981zk} is employed.
For the case of $^{76}$Ge and $^{82}$Se, the valence space consisting
on the 1$p_{3/2}$, 0$f_{5/2}$, 1$p_{1/2}$ and 0$g_{9/2}$ orbits is diagonalized
using the GCN28.50 interaction \cite{Menendez:2009dis}. Finally the 0$g_{7/2}$, 1$d_{3/2}$,
1$d_{5/2}$, 2$s_{1/2}$ and 0$h_{11/2}$ valence space and the GCN50.82
interaction  \cite{Menendez:2009dis} are used in the decays of $^{124}$Sn, $^{130}$Te and
$^{136}$Xe.
The results of the computation of the NME as a function of the neutrino mass are shown in \fig~\ref{fig:NME},
which is in agreement with the findings of \Refs~\cite{Doi:1981mj} and \cite{Benes:2005hn}.
As can be seen from this figure, the dependence of the matrix element on the nuclei is mild and all the curves show a similar behaviour.
We will discuss qualitatively the observed dependence of the NME on the neutrino mass as well as the phenomenology associated to the different mass
regimes in the next sections.

\subsection{Estimate of the uncertainties of the NME within the ISM \label{Sec:error}}

As we have pointed out several times throughout the present section,
since the computation of the NMEs can be rather challenging, a number of approximations need to be made.
In consequence, the results obtained will have some uncertainties, which we will now estimate.

First of all, when dealing with the ISM, we have to worry about the valence space and
the effective interaction used to obtain the nuclear wavefunctions.
The effect of having larger valence spaces was analyzed in Ref. \cite{Caurier:2007qn},
with the result that the NMEs increased, within a conservative analysis, by $\sim 15-20~\%$.
The same number was obtained in a QRPA calculation when it was quantified the effect of the orbits absent in a ISM valence space \cite{Simkovic:2008cu}.
Of the three valence spaces employed in the present work (see \Sec~\ref{Sec:wf}), only that corresponding to the $^{48}$Ca decay was not explored in Ref. \cite{Caurier:2007qn}, 
so that we can consider the former $\sim 15-20~\%$ increase as a general estimate of the uncertainty due to the configuration space.
As for the nuclear interaction, a moderate dependence
of $\sim 5-10~\%$ was found both in \Refs~\cite{Caurier:2007qn} and \cite{Menendez:2009oc}.
Since they study nuclei in different regions and we can consider all the effective
interactions employed of similar accuracy, we will also take this figure as general.

Until recently, another considerable source of uncertainty were the SRC.
However, it now seems that their contribution is rather small and that proper UCOM or Jastrow-type parametrizations can take these terms very well into
account, with a precision of $\sim 5~\%$~\cite{Engel:2009ha,Simkovic:2009pp}.
For its part, the variation in the NME due to different but reasonable values of the cutoffs appearing in the
FNS terms \cite{Towner:1995za,Kuzmin:2007kr} is also very small, less than 5~\%.

In addition, due to the fact that the NMEs have been calculated using the closure approximation,
we have to include an additional error of $5-10~\%$ to our results, as suggested by QRPA calculations \cite{Muto:1994hi}.
This is in agreement with the very soft variation that is seen in the NME as the parameter $\mu$ is modified.
Furthermore, we quantify the possible effect of missing terms in the operator
(next order terms in the current, nucleon recoil, $p$-wave emitted electrons)
by an additional uncertainty of less than 5~\% in our results.

It is currently under discussion whether the axial coupling should be quenched or not~\cite{Rodin:2006yk, Menendez:2009oc, Horoi:2009gz}.
In the ISM calculations presented here we take $g_{A}(0)=1.25$, \ie, we do not quench it, contrary to what is required by the single $\beta$ and two-neutrino $\beta\beta$ decays,
where a pure Gamow-Teller operator appears and its value has to be quenched to $g_{A}(0)=1.00$.
However, in the \znbb\ decay case the operator is more involved due to the extra radial dependence introduced by the virtual neutrino.
Moreover, in the case of the pure Gamow-Teller $J^P=1^+$ channel, it is not dominant in the \znbb\ process, and depending on its relative sign, quenching it may result even in an enhancement of the NME.
Until this issue is explored in more detail, we will take the most conservative option, allowing for a full quenching of the axial coupling and also for the quenching only of the $J^P=1^+$ channel.
Under these assumptions, taking into account that the Fermi part of the NME accounts for 10-15\% of the full NME in our ISM calculations and is never quenched,
we estimate a $\sim ^{+5~\%}_{-30~\%}$ error due to this effect.

Notice that the uncertainty in the valence space only moves the estimate up, the effect of axial quenching essentially moves it down,
while the remaining contributions are expected to be Gaussian-distributed.
Even though some of these errors may be correlated in a rather complicated way, as a first approximation we will take them as independent.
Altogether, adding every contribution in quadrature we expect an overall uncertainty in the final NME
of $\sim ^{+25~\%}_{-35~\%}$.

The above analysis applies to the case of light neutrino exchange.
For heavy neutrinos, the NMEs get very dependent on the SRC and FNS treatments.
In this case, a similar study to the one above gives a 15-20\% for SRC uncertainties and 10\% for FNS ones,
which would lead to a final $\sim ^{+35~\%}_{-40~\%}$ uncertainty.
However, one should take this number cautiously until the accuracy of the FNS approach for such heavy exchanged particles is firmly established \cite{Prezeau:2003xn}.

As an example, we will consider the case of the $^{76}$Ge decay. In \Ref~\cite{Menendez:2009oc}
the NME was obtained with different effective interactions and SRC, obtaining the interval $2.81 < M^{0\nu\beta\beta}(0) < 3.52$.
If we take into account the further uncertainties of the valence space, the FNS, the closure approximation, the next order hadronic current terms
and the $g_A(0)$ quenching, we end up with $2.11 < M^{0\nu\beta\beta}(0) < 3.98$.
This result will be used in \Sec~\ref{Sec:Models} to derive bounds on the neutrino masses from the \znbb\ decay process.

\section{General phenomenology}\label{sec:pheno}

According to \eq~(\ref{eq:t-1_rep}), the contribution of a single neutrino to the amplitude of \znbb\ decay is given by
\begin{equation}
 A_i \propto  m_i U_{ei}^2 M^{0\nu\beta\beta}(m_i),
\label{eq:amplitude1}
\end{equation}
where $m_i$ is the mass of the propagating neutrino and $M^{0\nu\beta\beta}(m_i)$
is the nuclear matrix element that characterizes the process and depends on the nucleus that undergoes the \znbb\ transition. 
Figure~\ref{fig:NME} shows two distinct regions where the behaviour of the NME as a function of the neutrino mass changes from almost constant
up to $m_i \simeq 100$~MeV to decreasing quadratically as the neutrino mass increases beyond $100$~MeV.
This behaviour is easily understood: the neutrino can be characterized as \emph{light} if $m_i^2 \ll |p^2|$ or \emph{heavy} if $m_i^2 \gg |p^2|$,
which would mean that the neutrino propagator in the NME would be dominated by $p^2$ or $m_i^2$, respectively, where $p$ is the momentum exchanged in the process.
As already mentioned in Sec.~\ref{Sec:NME}, in the \znbb\ decay $p^2 \simeq - \mathbf{p}^2 \simeq -(100{\rm~MeV})^2$.
We will therefore define two regimes:
\begin{itemize}
\item {\bf The light neutrino regime:} For $m_i \leq 100$~MeV, where the neutrino propagator is
\begin{equation}
\frac{1}{p^2-m_i^2} = \frac{1}{p^2} + \frac{m_i^2}{p^4} +\mathcal{O}\left(\frac{m_i^4}{p^6} \right) 
\end{equation}
and hence, the NME is maximum in this regime and is almost independent of the neutrino mass: 
\begin{equation}
M^{0\nu\beta\beta}(m_i) = M^{0\nu\beta\beta}(0) \left[ 1 + \frac{m_i^2}{p^2} +\mathcal{O}\left(\frac{m_i^4}{p^4} \right) \right].
\label{eq:lightexp}
\end{equation}
\item {\bf The heavy neutrino regime:}
For $m_i \geq 100$~MeV where the NME decreases as $m_i^{-2}$ providing an extra suppression to its contribution to the \znbb{} decay amplitude because of the neutrino propagator:
\begin{equation}
\frac{1}{p^2-m_i^2} = -\frac{1}{m_i^2} +\mathcal{O}\left(\frac{p^2}{m_i^4} \right).
\label{eq:heavyexp} 
\end{equation}

\end{itemize}

In principle, one could expect a more involved behaviour or even a resonance if $p^2 \simeq m_i^2$.
However, since the \znbb{} transition does not occur through an $s$-channel type diagram, the characteristic momentum transfer has $p^2 < 0$ with $|p^2| \simeq (100\ {\rm MeV})^2$.
Thus, the nuclear matrix element does not exhibit a resonant mass and the transition between the light and heavy neutrino mass regimes is relatively smooth (\cf, \fig~\ref{fig:NME}).
We will therefore not define a third transition region between the heavy and light regimes, as is sometimes done in the literature, since there is no new phenomenology associated to it. 

The usual bound derived from the \znbb\ process in the literature is obtained summing over the active neutrinos, implicitly neglecting the contribution of extra degrees of freedom.
With this assumption the only contribution comes from the neutrinos in the light regime, with $m_i \ll p^2$, which results in
\begin{equation}
 A_{\znbbeq} = \sum_{i=1}^3 A_i \propto  \sum_{i=1}^3 m_i U_{ei}^2 M^{0\nu\beta\beta}(m_i)
 =  M^{0\nu\beta\beta}(0) \sum_{i=1}^3 m_i U_{ei}^2 + \mathcal{O}\left(\frac{m_i^2}{p^2}\right),
\end{equation}
where $\sum_i m_i U_{ei}^2$ is the well-known expression used for the ``effective \znbb\ decay neutrino mass'':
\begin{equation}
\label{eq:mbb}
m_{\beta\beta} = m_1 c_{12}^2 c_{13}^2 + m_2 s_{12}^2 c_{13}^2 e^{2i\alpha_1} + m_3 s_{13}^2 e^{2i\alpha_2} ,
\end{equation} 
where $m_i$ are the masses of the neutrino mass eigenstates,
$c_{ij}=\cos \theta_{ij}$, $s_{ij}=\sin \theta_{ij}$, $\theta_{ij}$ are the neutrino mixing angles and $\alpha_i$ are combinations of the Majorana and Dirac phases.
It is important to note that this expression holds only when the SM neutrinos dominate the \znbb{} process.
However, if the SM is not extended, the Majorana mass required for the \znbb{} transition is forbidden. As we will discuss,
all extensions of the SM that induce a Majorana mass for the SM neutrinos imply the inclusion of extra degrees of freedom
that can contribute to the \znbb{} process and should be added to the SM decay amplitude. 

\begin{figure}
 \begin{center}
 \includegraphics[width=0.7\textwidth]{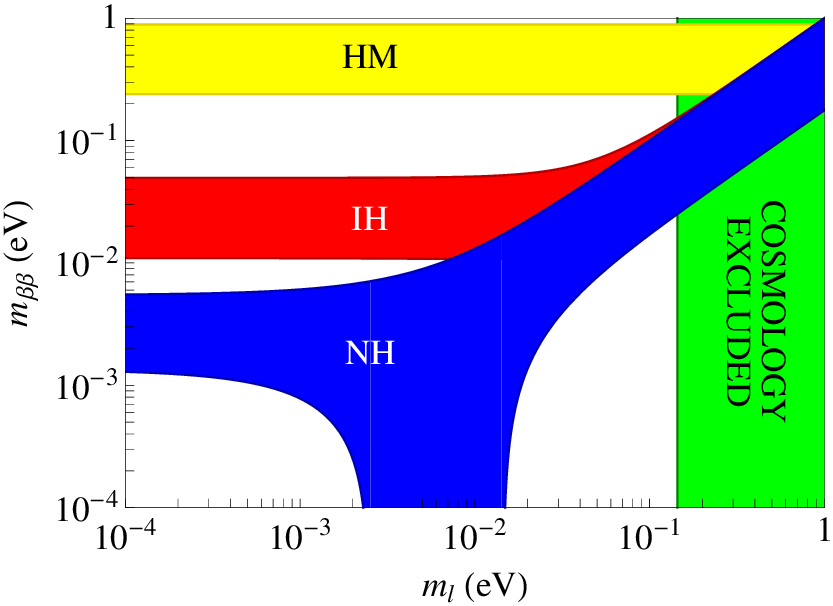}
 \bigskip 
  \caption{\label{fig:nh-ih} Allowed regions by neutrino oscillation data at $2 \sigma$ for $m_{\beta\beta}$ as a function of the mass of the lightest neutrino $m_l$, for normal (blue) and inverted (red) neutrino mass hierarchies.
The cosmologically disfavoured neutrino masses from cosmology are also shown (green band) as well as the Heidelberg-Moscow claim for \znbb{} decay (yellow band).}
 \end{center}
\end{figure}

Under the assumption that only the light neutrinos give a significant contribution to this process,
\eq~(\ref{eq:mbb}) can be combined with the present constraints on neutrino masses and mixings \cite{GonzalezGarcia:2010er} to derive \fig~\ref{fig:nh-ih},
where the allowed value of $m_{\beta\beta}$ as a function of the mass of the lightest neutrino $m_l$ is shown. However, as will be discussed in \Sec~\ref{Sec:Models}, this assumption is not always valid and \fig~\ref{fig:nh-ih} does not always provide an accurate description of the \znbb\ decay. In the same figure we also show the Heidelberg-Moscow claim for \znbb{} decay and the bound on the mass of the lightest neutrino from cosmology arguments~\cite{Hannestad:2010yi,Komatsu:2010fb}. Notice that the bounds from cosmology apply to the SM active neutrinos only. For extra sterile neutrinos, the cosmology bounds would depend on their mixing with the active ones, which would determine their abundance. Indeed, sterile neutrinos with masses $\sim 1$ keV larger than the cosmology bound are actually considered good candidates for warm dark matter \cite{Peebles:1982ib,Olive:1981ak,Dodelson:1993je}. As can be seen from the figure, there is a tension between the Heidelberg-Moscow claim, the contribution to \znbb{} 
from SM neutrinos and the present bounds on their mass from cosmology. In \Secs~\ref{sec:heavylight} and~\ref{sec:mix} we will discuss possible solutions to this tension.

\section{Applications to specific models of neutrino masses \label{Sec:Models}}

In this section we analyze the contributions to \znbb{} decay of the different mechanisms that lead to Majorana neutrino masses and can therefore induce the required lepton number violation for the process.
We will discuss here the tree-level realizations of the Weinberg $d=5$ effective operator:
\begin{eqnarray}
\label{eq:d=5}
\frac{c_{\alpha\beta}}{\Lambda} \left( \overline{L^c}_{\alpha} \tilde \phi^* \right) \left( \tilde \phi^\dagger \, L_{ \beta} \right) + \hc \;.
\end{eqnarray}
Here, $\phi$ denotes the SM Higgs field, which breaks the electroweak (EW) symmetry after acquiring its vacuum expectation value (vev) $v$,
$\Lambda$ is the scale of new physics that gives rise to the operator and we have used the definition $\tilde \phi = i \tau_2 \phi^*$.
This is the only $d=5$ operator that can be built from the SM particle content respecting both gauge and Lorentz invariance~\cite{Weinberg:1979sa}.
Since the low-energy effects of physics beyond the SM can be encoded in an expansion of effective operators of $d>4$ suppressed by inverse powers of $\Lambda^{d-4}$,
it is a natural expectation that this sole $d=5$ operator will be the least suppressed one.
It is then very suggestive that one of the few evidences we have for physics beyond the SM is the existence of small, but non vanishing, neutrino masses.
Indeed, after the Higgs field develops its vev, the operator of \eq~(\ref{eq:d=5}) induces a Majorana mass term for the SM neutrinos
$c_{\alpha\beta}(v^2/\Lambda) \overline{\nu_{\alpha L}^c} \nu_{\beta L}$, suppressed by the scale $\Lambda$. 

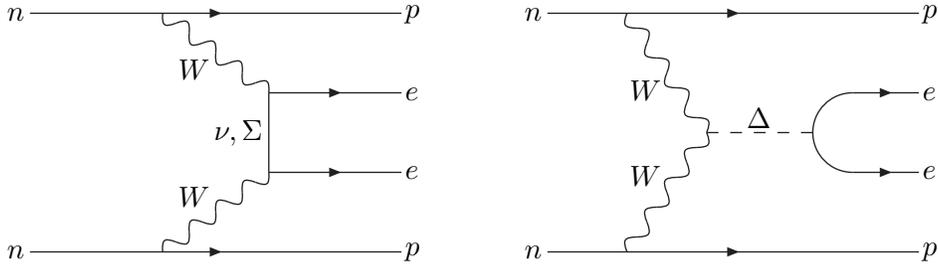
\begin{figure}
\begin{center}
%
% Neutrino diagram
\begin{picture}(160,110)(0,0)
 % Lower nucleon
 \Text(8,10)[r]{$n$}
 \Text(152,10)[l]{$p$}
 \ArrowLine(10,10)(150,10)
 % Upper nucleon
 \Text(8,100)[r]{$n$}
 \Text(152,100)[l]{$p$}
 \ArrowLine(10,100)(150,100)
 % Lower W
 \Text(78,27)[br]{$W$}
 \Photon(60,10)(100,40){3}{4}
 % Upper W
 \Text(78,83)[tr]{$W$}
 \Photon(60,100)(100,70){-3}{4}
 % Neutrino
 \Text(98,55)[r]{$\nu,\Sigma$}
 \Line(100,40)(100,70)
 % Lower electron
 \Text(152,40)[l]{$e$}
 \ArrowLine(100,40)(150,40)
 % Upper electron
 \Text(152,70)[l]{$e$}
 \ArrowLine(100,70)(150,70)
\end{picture}
\hspace{1cm}
%
%
%
% Type-II diagram
\begin{picture}(160,110)(0,0)
 % Lower nucleon
 \Text(8,10)[r]{$n$}
 \Text(152,10)[l]{$p$}
 \ArrowLine(10,10)(150,10)
 % Upper nucleon
 \Text(8,100)[r]{$n$}
 \Text(152,100)[l]{$p$}
 \ArrowLine(10,100)(150,100)
 % Lower W
 \Text(53,35)[br]{$W$}
 \Photon(40,10)(70,55){3}{4}
 % Upper W
 \Text(53,75)[tr]{$W$}
 \Photon(40,100)(70,55){-3}{4}
 % Doublet
 \Text(90,57)[b]{$\Delta$}
 \DashLine(70,55)(110,55){5}
 % Electron arc
 \CArc(125,55)(15,90,270)
 % Lower electron
 \Text(152,40)[l]{$e$}
 \ArrowLine(125,40)(150,40)
 % Upper electron
 \Text(152,70)[l]{$e$}
 \ArrowLine(125,70)(150,70)
\end{picture}
\end{center}
\caption{Feynman diagrams contributing to the \znbb{} transition rate from the exchange of fermions (left) and scalar triplets (right).}
\label{fig:diag}
\end{figure}

There are three different extensions of the SM particle content that lead to the operator of \eq~(\ref{eq:d=5}) after the extra mediators have been integrated out. They are known as ``seesaw'' mechanisms of \type{I}~\cite{Minkowski:1977sc,Yanagida:1979as,Mohapatra:1979ia,GellMann:1980vs}, where the heavy particles are fermion singlets, \type{II}~\cite{Magg:1980ut,Schechter:1980gr,Wetterich:1981bx,Lazarides:1980nt,Mohapatra:1980yp}, where scalars triplets are included, and \type{III}~\cite{Foot:1988aq,Ma:1998dn,Ma:2002pf,Hambye:2003rt}, where the SM is extended by fermion triplets.
All these extra degrees of freedom, required to induce the Majorana nature of the SM neutrinos, can also contribute to the \znbb\ process. The contribution of neutral fermions, such as the singlets and triplets added in the \type{I} and III seesaws, to the \znbb{} decay rate is depicted on the left side of \fig~\ref{fig:diag};
in particular this includes that of the light active neutrinos.
The contribution of the scalar triplet of the \type{II} seesaw (see, \eg, \Ref~\cite{Petcov:2009zr}) is depicted on the right side of \fig~\ref{fig:diag}. In both diagrams the $W$ lines can also be exchanged for the physical singly-charged scalar present in the \type{II} seesaw.
In principle, the contributions of the light active neutrinos and those of the extra degrees of freedom that are introduced should be combined. This is especially so in the case of the \type{I} seesaw, since important cancellations \cite{Halprin:1983ez,Leung:1984vy,Bamert:1994qh} are present in certain regimes \cite{deGouvea:2006gz} that are missed if the constraints are placed separately. If they are assumed to be independent, an extra contribution to the neutrino mass beyond those extra states is implicitly assumed. Moreover, taking into account the relations between the high- and low-energy parameters 
allows the derivation of stronger bounds on the former through the active neutrino contribution. It is then important not to neglect it, since the naive constraints stemming directly from the contribution of the extra degrees of freedom are generally much weaker. 

\subsection{\Type{I} seesaw models}

In this section we will discuss the phenomenology of \znbb{} decay when extending the Standard Model with fermion gauge singlets, \ie,  right handed neutrinos $\nu_{si}$.
The Standard Model Lagrangian is then extended as\footnote{Here, as in the following, we do not write out the kinetic terms of the new fields, which are assumed to be of the canonical form.}
\begin{eqnarray}\label{eq:seesawI}
\mathscr{L} &=& \mathscr{L}_\mathrm{SM} -\frac{1}{2} \overline{\nu_{si}} (M_N)_{ij} \nu_{sj}^{c} -(Y_{N})_{i\alpha}\overline{\nu_{si}} \widetilde \phi^\dagger
L_\alpha +\hc\; .
\end{eqnarray}
After the Higgs develops its vev the neutrino mass matrix is
\begin{equation}
M_\nu =
\left(
\begin{array}{cc}
0 & Y_N v/\sqrt{2} \\ Y_N^{T}v/\sqrt{2} & M_N
\end{array}
\right) .
\label{eq:massmat}
\end{equation}
This mass matrix can be diagonalized by a unitary mixing matrix $U$:
\begin{equation}
\label{diag}
U^* \diag\left\lbrace m_1, m_2,...,m_n \right\rbrace U^\dagger = M_\nu\,.
\end{equation}
We therefore have $n$ neutrino mass eigenstates with masses $m_i$ and mixings $U_{ei}$ with the electron.
Out of these, at least three mass eigenstates must be very light and form the main components of the active neutrinos, whose number is measured by the invisible decay width of the $Z$ \cite{Amsler:2008zzb}.
On the other hand, the masses of the extra states are not determined.
We thus have two contributions to the amplitude of \znbb{} decay, one from the light active neutrinos and another from the extra degrees of freedom:
\begin{equation}
 A \propto \sum_{i} m_i U_{ei}^2 M^{0\nu\beta\beta}(m_i) + \sum_{I} m_I U_{eI}^2 M^{0\nu\beta\beta}(m_I) ,
\end{equation}
where we have used capital letters to denote the mass index of the mostly sterile states and lowercase letters for that of the  mostly active states.
Depending on whether the extra mass eigenstates fall in the light or heavy neutrino mass regimes we can further split their respective contributions to the amplitude:
\begin{equation}
 A \propto \sum_i^{\rm light} m_i U_{ei}^2 M^{0\nu\beta\beta}(m_i) + \sum_I^{\rm light} m_I U_{eI}^2 M^{0\nu\beta\beta}(m_I) + \sum_I^{\rm heavy} m_I U_{eI}^2 M^{0\nu\beta\beta}(m_I) .
\label{eq:typeIampl}
\end{equation}
We can now distinguish three cases exhibiting very different phenomenologies depending on the mass regime of the extra mass eigenstates:

\subsubsection{All extra mass states in the light regime}

In this scenario all the mass eigenstates are lighter than $100$~MeV. In principle, this does not allow to explain the smallness of neutrino masses through the naive seesaw mechanism, \ie, with $\mathcal{O}\left(1 \right)$ Yukawa couplings. However, since the value of the parameter $M_N$ in \eq~(\ref{eq:seesawI}) is not restricted and it is technically natural for it to be small (for vanishing Majorana mass term the $B-L$ symmetry is recovered),
we believe this is a possibility worth exploring even if less appealing than the canonical \type{I} seesaw scenario.

\begin{figure}
\begin{center}
\includegraphics[width=0.7\textwidth]{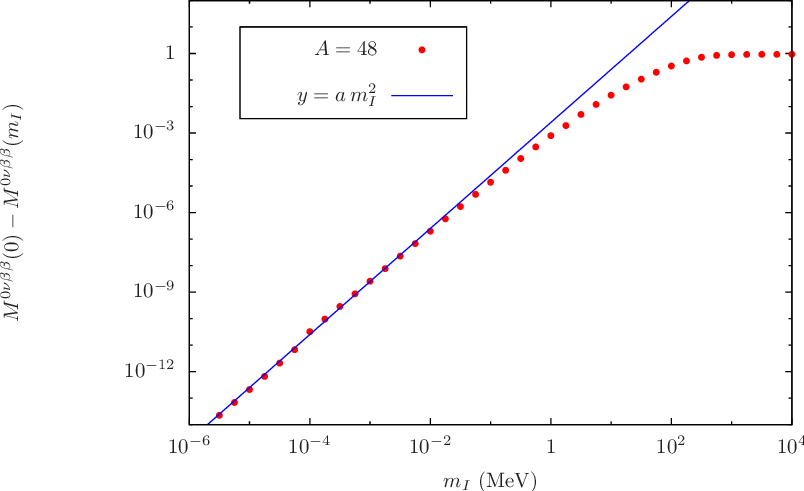}
 \bigskip 
 \caption{Nuclear matrix element cancellation in the light mass regime.
The example is taken for the $^{48}$Ca decay, but the same quadratic dependence is seen for the remaining nuclei studied.
The value of the parameter is $a=2.51\cdot10^{-3}$~MeV$^{-2}$. \label{fig:GIM}}
\end{center}
\end{figure}

Notice that, if all neutrinos belong the the light regime, \eq~(\ref{diag}) implies
\begin{equation}
\sum_i^{\rm light} m_i U_{ei}^2 + \sum_I^{\rm light} m_I U_{eI}^2 = 0 ,
\label{eq:GIMII}
\end{equation}
since the left-left entry of the mass matrix in \eq~(\ref{eq:massmat}) vanishes (it is forbidden by the SM gauge symmetry). Thus
\begin{eqnarray}
 A &\propto&  \sum_i^{\rm light} m_i U_{ei}^2 M^{0\nu\beta\beta}(m_i) + \sum_I^{\rm light} m_I U_{eI}^2 M^{0\nu\beta\beta}(m_I) 
 \nonumber
 \\
&\approx& -\sum_I^{\rm light} m_I U_{eI}^2 \left(M^{0\nu\beta\beta}(0) - M^{0\nu\beta\beta}(m_I)  \right).
\label{eq:zero}
\end{eqnarray}
On the other hand, for neutrino masses in the light regime, the nuclear matrix elements are basically independent of the neutrino mass (see \Fig~\ref{fig:NME})
$M^{0\nu\beta\beta}(m_I) \simeq M^{0\nu\beta\beta}(m_i) \simeq M^{0\nu\beta\beta}(0)$ and therefore 
the rate of \znbb\ decay in \eq~(\ref{eq:zero}) is very suppressed when all mass eigenstates are lighter than $\sim 100$~MeV.
This works in a way similar to the GIM suppression in flavour violating processes~\cite{Glashow:1970gm}.
Indeed, a process converting from flavour $\alpha$ to flavour $\beta$ is GIM suppressed due to the unitarity relation $\sum_i U_{\alpha i}U^*_{\beta i} = 0$,
so that only the mass of the propagating particle does not make the cancellation exact and a suppression of $\Delta m^2/M_W^2$ is found,
where $\Delta m^2$ is the mass squared difference between the different propagating particles.
In our \znbb{} process we have \eq~(\ref{eq:GIMII}) operating a similar cancellation and again it is the different neutrino masses in the NME of \eq~(\ref{eq:zero}) that would prevent a full cancellation leading to a suppression of $\Delta m^2/p^2$ with $|p^2| \simeq (100\ {\rm MeV})^2$ [see the second term in \eq~(\ref{eq:lightexp}) and, \eg, \Ref~\cite{Pascoli:2007qh}].
The $\Delta m^2/p^2$ dependence of the $M^{0\nu\beta\beta}(0) - M^{0\nu\beta\beta}(m_I)$ term that drives the contribution to \znbb{} 
in \eq~(\ref{eq:zero}) is depicted in \fig~\ref{fig:GIM}.
As expected, the two contributions cancel up to a factor $m_I^2/p^2$ with $|p^2| \simeq (100\ {\rm MeV})^2$ and deviations from this behaviour start to be non negligible for $m_I \gtrsim 1$~MeV.

\begin{figure}
\begin{center}
\includegraphics[width=0.7\textwidth]{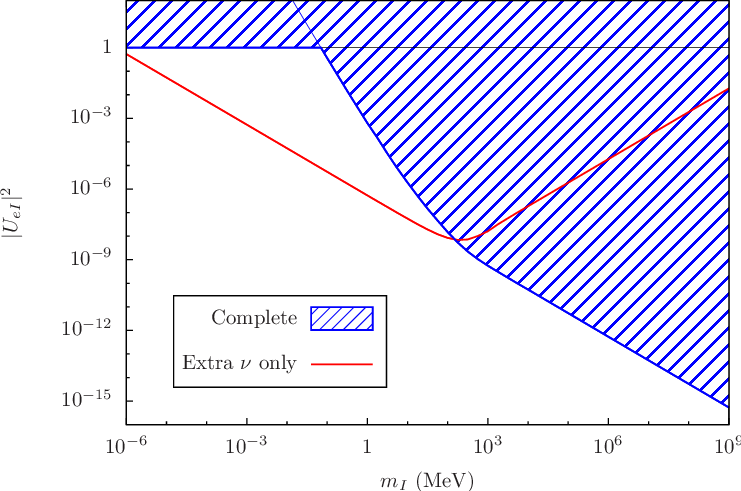}
 \bigskip 
 \caption{Bounds from CUORICINO on the extra neutrino mixing from \znbb\ decay in $^{130}$Te, with a 90~\%
CL half-life \cite{Arnaboldi:2008ds}. We have assumed the extra neutrinos are non hierarchical and show the bounds as a function of their 
common mass. We compare the case in which the contribution from the SM neutrinos is properly taken into account (striped area) to that in which only the extra contribution is considered (above the red line). 
\label{fig:U_bounds}}
\end{center}
\end{figure}

We want to emphasize that phenomenological analyses (see, \eg, \Refs~\cite{Benes:2005hn} and \cite{Atre:2009rg})
that use the non observation of \znbb{} decay to derive bounds on the mixing of an extra light neutrino with mass around $\sim 100$ MeV or below neglecting the contribution of SM neutrinos, implicitly assume that the cancellation described in \eq~(\ref{eq:zero}) does not take place. This would be the case if some extra contribution to the neutrino masses is present and \eq~(\ref{eq:GIMII}) is consequently modified. Examples of this situation, which imply some degree of cancellation between the different contributions to neutrino masses, will be discussed in \Secs~\ref{sec:heavylight} and~\ref{sec:mix}. In the absence of these extra contributions, the remaining leading term, GIM suppressed as $\Delta m^2/p^2$, can then be used to derive the corrected bound on the mixing of the extra state.
In \fig~\ref{fig:U_bounds} we show this bound using the constraints on \znbb\ decay in $^{130}$Te from CUORICINO, with a 90~\% CL half-life \cite{Arnaboldi:2008ds}
and compare it to the one that would be obtained neglecting this cancellation and considering only the extra neutrino contribution.
It is clearly seen that the bound indeed weakens when $m_I < 100$ MeV. In particular, for $m_I < 1$ MeV, the $M^{0\nu\beta\beta}(0) - M^{0\nu\beta\beta}(m_I)$ suppression in \fig~\ref{fig:GIM} becomes almost $10^{-3}$, so that the \znbb{} decay rate would be suppressed by six orders of magnitude and for $m_I < 100$ keV the bound on the mixing from \fig~\ref{fig:U_bounds} becomes meaningless since $\sum_I U^2_{eI} < 1$. Clearly, if the extra states that accommodate neutrino masses are all in this regime the \znbb{} decay becomes experimentally inaccessible even if neutrinos are Majorana particles. 
An important sub-case is when $M_N = 0$ with three extra neutrinos, which corresponds to the case of Dirac neutrinos.
In this scenario, the cancellation is perfect since the left and right handed states are exactly degenerate,
which corresponds to the fact that there is no lepton number violation for Dirac neutrinos.

In order to derive the constraints of \fig~\ref{fig:U_bounds} we have assumed as an example that all the extra states are degenerate in mass (or that there is only one extra state).
However, a similar GIM-like cancellation would also be present when assuming a different hierarchy as long as all the extra states are in the light regime.
The extra contribution in the light regime can only dominate over the light active one and avoid the GIM-like cancellation if the model is extended with other heavier states.
This situation also implies a certain degree of fine-tuning between the extra contributions and will be discussed in detail in \Secs~\ref{sec:heavylight} and \ref{sec:mix}. 

\subsubsection{All extra mass states in the heavy regime}

This is the simplest situation and the one implicitly assumed when using \eq~(\ref{eq:mbb}) or \fig~\ref{fig:nh-ih} to forecast the rate of \znbb\ decay.
It also corresponds to the canonical implementation of the \type{I} seesaw mechanism,
where the Majorana mass of the right handed neutrinos in \eq~(\ref{eq:seesawI}) is assumed to be very large $M_N \gg Y_Nv/\sqrt{2}$
so that the smallness of the active neutrino masses is naturally explained and extra heavy eigenstates are obtained.
Since the extra states would fall in the heavy neutrino mass regime their nuclear matrix elements are very suppressed:
$M^{0\nu\beta\beta}(m_I) \ll M^{0\nu\beta\beta}(m_i)$ (see \fig~\ref{fig:NME}).
Furthermore, \eq~(\ref{diag}) implies that
\begin{eqnarray}
\sum_i^{\rm light} m_i U_{ei}^2 + \sum_I^{\rm heavy} m_I U_{eI}^2 = 0 ,
\label{eq:GIMI}
\end{eqnarray}
so that the \znbb{} decay amplitude is

\begin{eqnarray}
 A &\propto&  \sum_i^{\rm light} m_i U_{ei}^2 M^{0\nu\beta\beta}(m_i) + \sum_I^{\rm heavy} m_I U_{eI}^2 M^{0\nu\beta\beta}(m_I) 
 \nonumber
 \\
\nonumber
&\approx& -\sum_I^{\rm heavy} m_I U_{eI}^2 \left(M^{0\nu\beta\beta}(0) - M^{0\nu\beta\beta}(m_I)  \right) \\
&\approx& -\sum_I^{\rm heavy} m_I U_{eI}^2 M^{0\nu\beta\beta}(0) 
= \sum_i^{\rm light} m_i U_{ei}^2 M^{0\nu\beta\beta}(0).
\label{eq:heavyfromlight}
\end{eqnarray}
The contribution from the light active neutrinos thus dominates and \fig~\ref{fig:nh-ih} provides an accurate prediction for the \znbb{} transition rate. 

Notice that, using the contribution of the extra states $m_I U_{eI}^2 M^{0\nu\beta\beta}(m_I)$ to the \znbb{} process
in order to derive a bound on their mixing $U_{eI}$ would lead to rather weak constraints since it is very subleading (see, \eg, \Refs~\cite{Belanger:1995nh,Simkovic:1999re,delAguila:2008cj}).
On the other hand, \eq~(\ref{eq:heavyfromlight}) can be used to express the dominant light neutrino contribution as a function of the heavy parameters:
$-m_I U_{eI}^2 M^{0\nu\beta\beta}(0)$.
This expression then allows to derive a much stronger constraint on $U_{eI}$ \cite{Xing:2009ce}.
This constraint is also shown in \fig~\ref{fig:U_bounds} when $m_I > 100$~MeV and compared to that when only the contribution of the heavy neutrinos is considered.
In order to derive the constraints of \fig~\ref{fig:U_bounds} we have assumed as an example that all the extra states are degenerate in mass (or that there is only one extra state).
However, a similar behaviour is found when a different hierarchy is considered as long as all the extra states are in the heavy regime.

\subsubsection{Extra mass states in the light and heavy regimes}
\label{sec:heavylight}

In this scenario we would have the full contribution to \znbb{} decay of \eq~(\ref{eq:typeIampl}) and the constraint from \eq~(\ref{diag}):
\begin{equation}
\sum_i^{\rm light} m_i U_{ei}^2 + \sum_I^{\rm light} m_I U_{eI}^2 + \sum_I^{\rm heavy} m_I U_{eI}^2 = 0 .
\label{eq:GIMIII}
\end{equation}
As discussed above, for the neutrinos in the heavy regime, the NME receives an extra suppression to their contribution to the \znbb{} decay rate
and the leading terms stem from the light states:
\begin{equation}
 A \propto  \sum_i^{\rm light} m_i U_{ei}^2 M^{0\nu\beta\beta}(m_i) + \sum_I^{\rm light} m_I U_{eI}^2 M^{0\nu\beta\beta}(m_I) 
 \simeq - \sum_I^{\rm heavy} m_I U_{eI}^2 M^{0\nu\beta\beta}(0) .
\label{eq:tuning}
\end{equation}
However, in this case the GIM-like cancellation is prevented since the heavy contribution is suppressed. This scenario thus offers the richest phenomenology.
In particular, it is possible to satisfy \eq~(\ref{eq:GIMIII}) even in a situation where $m_i U_{ei}^2 \ll m_I U_{eI}^2$
by canceling the contribution of the extra heavy states against that of the extra light ones while keeping the light neutrino masses small. 
This implies a certain level of fine-tuning since extra sterile neutrinos in both the heavy and light regimes are necessary and some degree of cancellation between their respective contributions is required in order to keep the neutrino masses small.
However, in such a situation, the contribution of the light extra states could dominate over that of the active and induce a rate for the \znbb{} process larger than the one forecasted from \fig~\ref{fig:nh-ih}.
This could thus be a possible solution to an eventual discrepancy between a positive result in \znbb{} decay and a negative result in the searches for neutrino masses in cosmology.
Indeed, the bounds from cosmology apply to the active SM neutrinos only. 

\begin{figure}
\begin{center}
\includegraphics[width=0.7\textwidth]{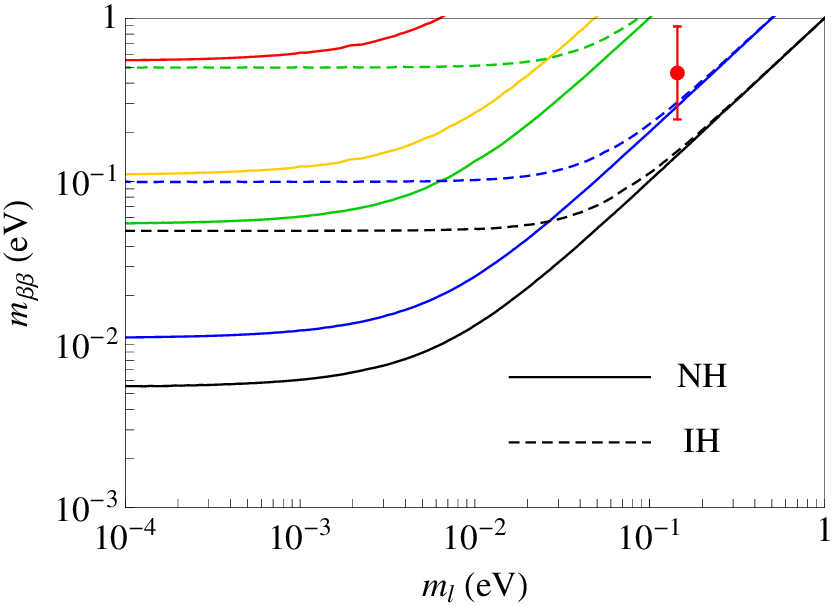}
 \bigskip 
 \caption{The level of cancellation between the different contributions to the neutrino mass necessary to reconcile an eventual discovery of \znbb\ decay
with size $m_{\beta \beta}$ with a bound on the lightest neutrino mass $m_l$ from future cosmology or $\beta$ decay experiments.
The solid and dashed lines are for normal and inverted mass hierarchy respectively.
The red, yellow, green, blue and black contours are for cancellations to the level of $1~\%$, $5~\%$, $10~\%$, $50~\%$ and no cancellation respectively.
The dot represents the present Heidelberg-Moscow claim for $m_{\beta \beta}$  and the cosmology bound on $m_l$
and requires a tuning of $\sim 50~\%$ in order to be explained through sterile neutrinos.
\label{fig:tuning}}
\end{center}
\end{figure}

As an example, we will here consider the Heidelberg-Moscow claim for a positive \znbb{} decay signal \cite{KlapdorKleingrothaus:2006ff}.
The accommodation of this signal through only SM neutrinos [see \eq~(\ref{eq:mbb})] would require $0.24~\mathrm{eV} < m_{\beta \beta} < 0.89~\mathrm{eV}$
at $2\sigma$, where the allowed numbers have been obtained with the ISM results of \Sec~\ref{Sec:error} following the rather conservative procedure described in Ref. \cite{Faessler:2008xj}.
Almost all the error bar comes from the theoretical error of the NME, which is much larger than the one associated to the experimental claim.
As can be seen in \fig~\ref{fig:nh-ih}, the interpretation of this claim as light active SM neutrinos is very disfavoured (see, \eg, \Ref~\cite{Fogli:2008ig})
by the constraints from cosmology and neutrino oscillation data.
However, this signal could be accommodated in a model with heavier neutrinos (which are not bounded by cosmology) mediating the process.
Indeed, following \eq~(\ref{eq:tuning}), we could reinterpret the result as 
\begin{equation}
0.24~\mathrm{eV} < \left| \sum_I^{\rm heavy} m_I U_{eI}^2 \right| < 0.89~\mathrm{eV}.
\end{equation}
This larger contribution to \znbb{} decay would not be in conflict with neutrino masses if the extra heavy and light neutrino contributions cancel each other in \eq~(\ref{eq:GIMIII}).
The level of cancellation required to accommodate the the Heidelberg-Moscow claim with sterile neutrinos is actually only at the $\sim 50~\%$ level,
since the light active neutrinos only fail to explain it by about a factor two, given the bounds we have on their masses from cosmology.
In \fig~\ref{fig:tuning} we show the degree of cancellation that would be required to accommodate an eventual stronger bound on the mass of the lightest neutrino $m_l$
from cosmology~\cite{Hannestad:2010yi,Komatsu:2010fb} with a discovery of \znbb\ decay requiring a given $m_{\beta \beta}$ when interpreted as the contribution of the SM neutrinos alone.
The contours for no tension between the mass bound and $m_{\beta\beta}$ are shown together with the situations that require cancellations at the level of $50~\%$, $10~\%$, $5~\%$ and $1~\%$.
The dot represents the present situation of the Heidelberg-Moscow claim that requires a mild $50~\%$ cancellation to avoid conflict with the cosmological bounds on $m_l$.
Notice that, since a minimum size of $m_{\beta \beta} \sim 10^{-2}$ eV is guaranteed for an inverted hierarchy, the required level of cancellation never exceeds the $10~\%$ in this situation.
On the other hand, if the neutrino mass hierarchy is found to be normal, tunings up to $\sim 1~\%$ would be necessary to reconcile a discovery of $m_{\beta \beta}$
on the same level as the present Heidelberg-Moscow claim with an eventual bound of $m_l < 10^{-2}$ eV.

\subsection{\Type{II} seesaw models}

In the \type{II} seesaw, the Standard Model content is expanded by the addition of a scalar $SU(2)$ triplet with hypercharge $2$ (where the hypercharge is defined such that $Q=Y/2+T_3$):
\begin{equation}
\Delta =
\left(
\begin{array}{cc}
\Delta^+/\sqrt{2} & \Delta^{++} \\ \Delta^0 & -\Delta^+/\sqrt{2}
\end{array}
\right) .
\end{equation}
The scalar triplet couples to the lepton doublet through the Yukawa terms:
\begin{equation}\label{eq:seesawII}
\mathscr{L} = \mathscr{L}_\mathrm{SM} -(Y_{\Delta})_{\alpha \beta} \overline{L^c}_\alpha i \tau_2 \Delta L_\beta +\hc ,
\end{equation}
where $\tau_2$ is the second Pauli matrix. Furthermore, the scalar triplet $\Delta$ has a coupling $\mu$ to a pair of Higgs fields, so that it gets an induced vev after EW symmetry breaking:
$v_\Delta = \mu v^2/2M^2_{\Delta}$, where $M_\Delta$ is the mass of the scalar triplet. The vev of the triplet then induces a Majorana neutrino mass
\begin{equation}
\label{eq:massII}
m^\Delta_\nu=2Y_{\Delta}v_\Delta = Y_{\Delta} \frac{\mu v^2}{M^2_{\Delta}}
\end{equation}
in \eq~(\ref{eq:seesawII}).
This in turn implies that, at low energies, \eq~(\ref{eq:GIMI}) is modified to
\begin{equation}
\sum_i^{\rm light} m_i U_{ei}^2 = \left( m^{\Delta}_\nu\right) _{ee},
\label{eq:noGIMII}
\end{equation}
making $m^\Delta_\nu$ the analogous contribution to the one of the heavy neutrinos in the \type{I} seesaw.

The \znbb{} process can be mediated both by the neutrinos and by the scalar triplet (see \fig~\ref{fig:diag}).
However, such charged scalars would have been produced at Tevatron for masses below 100~GeV~\cite{:2008iy}.
Thus, the contribution of the charged scalar is suppressed with respect to the neutrino one. The diagram in the right side of \Fig~\ref{fig:diag} 
replaces the neutrino propagator by a scalar propagator and thus its amplitude is suppressed by a factor
$\sim p^2/M_\Delta^2  < 10^{-6}$ with respect to the SM neutrino contribution. 
The diagram in the left side of \Fig~\ref{fig:diag} in which one of the $W$ bosons is replaced by the physical charged scalar 
are also suppressed. The scalar is an admixture of the charged components of the Higgs doublet and the scalar triplet and its coupling to the quarks is therefore proportional to the quark mass. Thus, the amplitude of these contributions turn out to be suppressed by a factor $\sim m_q/M_\Delta  < 10^{-5}$, where $m_q$ is the mass of either the up or down quark. Therefore, in this scenario, as in the \type{I} seesaw with all extra states heavy,
the light active neutrino contribution dominates and the usual description of \znbb{} decay in \fig~\ref{fig:nh-ih} applies.

\subsection{\Type{III} seesaw models}

In the \type{III} seesaw models the Standard Model is expanded by fermion $SU(2)$ triplets with zero hypercharge:
\begin{equation}
\Sigma =
\left(
\begin{array}{cc}
\Sigma^0/\sqrt{2} & \Sigma^{+} \\ \Sigma^- & -\Sigma^0/\sqrt{2}
\end{array}
\right) .
\end{equation}
The fermion triplets couple to the SM lepton doublets and the Higgs field through the Yukawa terms and have Majorana mass terms of their own:
\begin{eqnarray}
\label{eq:seesawIII}
\mathscr{L} &=& \mathscr{L}_\mathrm{SM} -\frac{1}{2} (M_{\Sigma})_{ij} \mathrm{Tr}\left( \overline{\Sigma}_i  \Sigma^{c}_j \right) -(Y_{\Sigma})_{i\alpha} \tilde{\phi}^\dagger \overline{\Sigma}_i i \tau_2 L_\alpha +\hc\; .
\end{eqnarray}
The \znbb\ decay phenomenology of the \type{III} seesaw is then completely analogous to that of the \type{I} with the neutral component of the triplet playing the role of the right handed neutrino,
except that, since the triplet also has charged components, stringent lower bounds on its mass exist and in practice only the heavy mass eigenstate regime is available.
The situation then reduces to the one for the \type{II} seesaw instead, \ie, the same \znbb\ phenomenology applies, with the replacement
\begin{equation}
\label{eq:massIII}
m_\nu^\Delta \longrightarrow m^\Sigma_\nu=\frac{v^2}{2}Y^T_{\Sigma}M^{-1}_\Sigma Y_\Sigma.
\end{equation}

\subsection{Mixed seesaw models}
\label{sec:mix}

It is interesting to note that the same phenomenology that could stem from a \type{I} seesaw with both heavy and light eigenstates
can also arise from a \type{II} or III seesaw in combination with \type{I} sterile neutrinos in the light regime.
Indeed, adding a \type{II} or III contribution to neutrino masses $m^{\Delta,\Sigma}$ as in \eqs~(\ref{eq:massII}) and (\ref{eq:massIII}),
\eq~(\ref{eq:massmat}) would instead read
\begin{equation}
M_\nu =
\left(
\begin{array}{cc}
m^{\Delta,\Sigma} & Y_N v/\sqrt{2} \\ Y_N^{T}v/\sqrt{2} & M_N
\end{array}
\right) .
\label{eq:massmat2}
\end{equation}
This in turn implies that \eq~(\ref{eq:GIMII}) is modified to
\begin{equation}
\sum_i^{\rm light} m_i U_{ei}^2 + \sum_I^{\rm light} m_I U_{eI}^2 = m^{\Delta,\Sigma}_{ee} .
\label{eq:noGIM}
\end{equation}
Thus, it is possible to have a dominant contribution to \znbb{} decay from the extra light sterile neutrinos if $m_I U_{eI}^2 \gg m_i U_{ei}^2$,
while \eq~(\ref{eq:noGIM}) and the smallness of neutrino masses is respected by a cancellation between $m_I U_{eI}^2$ 
and $m^{\Delta,\Sigma}_{ee}$. The level of the cancellation required also corresponds to the one depicted in \fig~\ref{fig:tuning}.
Indeed, the amplitude of \znbb{} decay would now be 
\begin{equation}
 A \propto  \sum_i^{\rm light} m_i U_{ei}^2 M^{0\nu\beta\beta}(m_i) + \sum_I^{\rm light} m_I U_{eI}^2 M^{0\nu\beta\beta}(m_I) 
 \simeq  m_{ee}^{\Delta,\Sigma} M^{0\nu\beta\beta}(0),
\label{eq:tuning2}
\end{equation}
and, as an example, the Heidelberg-Moscow claim can be interpreted as
\begin{equation}
0.24~\mathrm{eV} < \left| m_{ee}^{\Delta,\Sigma} \right| < 0.89~\mathrm{eV}
\end{equation}
in this context.

\section{Summary and conclusions}
\label{sec:summary}

We have discussed the general phenomenology of neutrinoless double beta decay (\znbb\ decay) in different types of seesaw models.
In particular, we have focused on the contributions of the extra degrees of freedom in different mass regimes without assuming preference to a particular mass scale.
In order to do this, we computed the nuclear matrix element (NME) involved in the decay amplitudes as function of the mass of the mediating field,
detailing all the assumptions performed in each step and estimating the final error due to the approximations taken to be at most 30~\% for light neutrinos and around 40~\% for heavy neutrinos.
The results of this computation are publicly available in Appendix~\ref{nme_table}. %at \Ref~\cite{webpage}.
In particular, the behaviour of the NME is found to be that which can be expected from the propagator $1/(p^2-m^2)$, where $p^2 \sim -(100$~MeV$)^2$ is the typical momentum transfer between the nucleons.
Thus, the NMEs are essentially constant for $m < 100$~MeV (light regime) and decrease as $m^{-2}$ for $m > 100$~MeV (heavy regime).
The transition region around 100~MeV is smooth and no significantly new phenomenology takes place at this regime.

In our discussion we have seen that, for the \type{I} seesaw, a number of possibilities exist.
In the case where all the masses of the extra fermion singlets are in the heavy regime,
the contribution of these states to \znbb\ decay is negligible and difficult to constrain directly.
However, the contribution from the light left handed neutrinos
can be rewritten in terms of the masses and mixings of the heavy states, which results in quite stringent bounds on their mixings. 
On the contrary, if the extra fermion singlet states are instead in the light regime,
 a GIM-like cancellation with the left handed neutrino contributions occurs.
In this situation, neutrinos are Majorana particles, but the \znbb\ decay rate becomes unobservable, being suppressed by at least six orders of magnitude if the mass of the extra states is below 1~MeV.
Thus, considering the extra states alone and neglecting the SM neutrino contribution will result in bounds on their mixings which are generally too strong.
Such bounds only apply when considering extra contributions to neutrino masses beyond the light extra states in order to prevent the cancellation from taking place.
Dirac neutrinos, where the Majorana mass term of the extra fermion states is zero, are a special case of the situation with only light extra states.
In this scenario, the right and left handed states are exactly degenerate and the cancellation is exact.

If there are extra states in both the light and heavy regimes, then the main contribution to the \znbb\ transition
could come from the light extra states, although some fine-tuning is necessary.
As such, this could be a way to reconcile a large \znbb\ decay rate (\eg, the Heidelberg-Moscow claim) with more stringent cosmological bounds
with a mild cancellation of about 50~\%.

As for the other types of seesaws, current bounds from accelerator experiments place the extra degrees of freedom in the heavy regime.
This effectively reduces the situation to that which appears for the \type{I} seesaw with only heavy extra states.
However, in mixed seesaw models, the situation can instead resemble that of the \type{I} with states in both regimes
and thus be used to reconcile large \znbb\ decay rates with cosmological bounds.

In conclusion, the contribution to \znbb\ decay from the light active neutrinos can be forecasted by combining present and future neutrino oscillation data
on the neutrino mixing and mass hierarchy with probes of the absolute neutrino mass scale such as cosmology.
These predictions can be compared to future \znbb\ decay searches so as to gain information on the origin and nature of the neutrino masses.
In this comparison, we can distinguish the following scenarios:

\begin{itemize}

\item {\bf The $\bld\znbbeq$ process is observed to be in agreement with the forecasted rates.}
This indicates that the light active neutrinos dominate the \znbb\ decay rate.
Since new degrees of freedom are in any event required to give the light neutrinos Majorana masses,
this implies that there is necessarily new physics above the nuclear scale, so that its contribution is suppressed. 

\item {\bf The $\bld\znbbeq$ process is observed to be smaller than the forecasted rates.}   
This means that there is a partial cancellation between the active and extra neutrino contributions.
Sterile neutrinos around the nuclear scale are then necessary.
A higher mass would imply too big a suppression through their NME to show any sizable cancellation,
while too small masses would make the GIM-like cancellation exact.

\item {\bf The $\bld\znbbeq$ process is observed to be larger than the forecasted rates.}
In this situation the light active neutrinos cannot dominate the \znbb\ decay rate.
Extra sterile neutrinos, lighter or around the nuclear scale, could have a significant contribution and reconcile the observations.
However, the GIM-like cancellation between both contributions has to be avoided.
This implies either a cancellation between extra neutrinos both above and below the nuclear scale (see \Sec~\ref{sec:heavylight})
or between the extra neutrinos and a type-II or III seesaw contribution (see \Sec~\ref{sec:mix}).
This is the case that would correspond to a confirmation of the Heidelberg-Moscow claim.

\item{\bf The $\bld\znbbeq$ process is not observed but was forecasted.} 
While this could imply that neutrinos are Dirac and not Majorana particles, it can also be the case that neutrinos are Majorana
but extra sterile neutrinos below the nuclear scale are present. Thus, the GIM-like cancellation takes place and the \znbb\ decay rate becomes unobservable.

\item{\bf The $\bld\znbbeq$ process is not observed and was not forecasted.} 
This is the most pessimistic scenario since it is impossible to draw any conclusion on the nature and origin of neutrino masses. 

\end{itemize}

\begin{acknowledgments}

We are specially indebted to Andrea Donini for carefully reading through this paper and providing valuable comments.
We also acknowledge very interesting and fruitful discussions with Carla Biggio, Belen Gavela, Pilar Hernandez, Alessandro Mirizzi, Ann Nelson, Alfredo Poves and Achim Schwenk.

This work was supported by
the European Union through the European Commission Marie Curie Actions Framework Programme 7 Intra-European Fellowship: Neutrino Evolution [M.B.]
and the European Commission Framework Programme 07 Design Study EURO$\nu$, project 212372 [J.L.P];
the Spanish Ministry of Education and Science through the CUP Consolider-Ingenio 2010, project CSD2008-0037 [J.L.P.] and grants FPA2009-09017 [J.L.P] and FPA2009-13377 [J.M.];
the DFG through cluster of excellence ``Origin and Structure of the Universe'' [E.F.M.] and grant SFB 634 [J.M.];
the Helmholtz Association through the Helmholtz Alliance Program, contract HA216/EMMI ``Extremes of Density and Temperature: Cosmic Matter in the Laboratory'' [J.M.];
and the Comunidad Aut\'onoma de Madrid through project HEPHACOS-CM (S2009ESP-1473) [J.L.P., J.M.]. 

\end{acknowledgments}

\newpage

\appendix

\section{Nuclear matrix elements\label{nme_table}}

\begin{table}[h!]
\caption{Nuclear matrix elements for different neutrino masses for the \znbb\ decays of $^{48}$Ca, $^{76}$Ge, $^{82}$Se, $^{124}$Sn, $^{130}$Te and $^{136}$Xe.
The calculations include UCOM short-range correlations and unquenched axial coupling $g_A=1.25$.}
%\begin{center}
\smallskip
%@{\extracolsep{\fill}}
\noindent\makebox[\textwidth]{%
\begin{tabular}{c|c|c|c|c|c|c}
\hline
\hline
$m_{\nu}$~(MeV) & $^{48}$Ca & $^{76}$Ge & $^{82}$Se & $^{124}$Sn & $^{130}$Te & $^{136}$Xe \\
\hline
     $1.0\times10^{-9}$  &   0.938      &     2.79      &     2.61     &      2.79      &     2.60     &      2.15 \\
     $1.0\times10^{-6}$  &   0.938      &     2.79      &     2.61     &      2.79      &     2.60     &      2.15 \\
     $1.0\times10^{-3}$  &   0.938      &     2.79      &     2.61     &      2.79      &     2.60     &      2.15 \\
      1.000      &   0.938      &     2.79      &     2.61     &      2.78      &     2.60     &      2.15 \\
      1.778      &   0.936      &     2.78      &     2.60     &      2.78      &     2.59     &      2.15 \\
      3.162      &   0.933      &     2.77      &     2.59     &      2.77      &     2.59     &      2.14 \\
      5.623      &   0.926      &     2.74      &     2.56     &      2.74      &     2.57     &      2.12 \\
      10.00      &   0.912      &     2.68      &     2.50     &      2.69      &     2.53     &      2.08 \\
      17.78      &   0.883      &     2.55      &     2.39     &      2.58      &     2.44     &      2.01 \\
      31.62      &   0.830      &     2.34      &     2.19     &      2.38      &     2.28     &      1.87 \\
      56.23      &   0.741      &     2.01      &     1.88     &      2.06      &     1.99     &      1.63 \\
      100.0      &   0.600      &     1.55      &     1.44     &      1.59      &     1.57     &      1.28 \\
      177.8      &   0.410      &     1.01      &    0.940     &      1.04      &     1.04     &     0.846 \\
      316.2      &   0.219      &    0.522      &    0.485     &     0.542      &    0.549     &     0.442 \\
      562.3      &   0.0887  &    0.208      &    0.193     &     0.218      &    0.222     &     0.178 \\
      1000      &   0.0311  &    0.0729  &    0.0673 &     0.0769  &    0.0785 &     0.0627 \\
      1778      &   8.94$\times10^{-3}$  &    0.0209  &    0.0193 &     0.0222  &    0.0227 &     0.0181 \\
      3162      &   2.75$\times10^{-3}$  &    6.46$\times10^{-3}$  &    5.96$\times10^{-3}$ &     6.85$\times10^{-3}$  &    7.01$\times10^{-3}$ &     5.58$\times10^{-3}$ \\
      5623      &   8.61$\times10^{-4}$  &    2.02$\times10^{-3}$  &    1.86$\times10^{-3}$ &     2.15$\times10^{-3}$  &    2.19$\times10^{-3}$ &     1.75$\times10^{-3}$ \\
     $1.0\times10^{4}$  &   2.71$\times10^{-4}$  &    6.37$\times10^{-4}$  &    5.88$\times10^{-4}$ &     6.77$\times10^{-4}$  &    6.92$\times10^{-4}$ &     5.51$\times10^{-4}$ \\
     $1.0\times10^{6}$  &   2.71$\times10^{-8}$  &    6.36$\times10^{-8}$  &    5.87$\times10^{-8}$ &     6.76$\times10^{-8}$  &    6.91$\times10^{-8}$ &     5.51$\times10^{-8}$ \\
     $1.0\times10^{9}$  &   2.71$\times10^{-14}$  &   6.36$\times10^{-14}$  &   5.87$\times10^{-14}$ &    6.76$\times10^{-14}$  &   6.91$\times10^{-14}$ &    5.51$\times10^{-14}$ \\
\hline
\hline
\end{tabular}}
\end{table}

\end{document}